\documentclass{mem}
\usepackage{natbib}
\usepackage{txfonts}
\usepackage{balance}
\usepackage{graphicx}
\idline{75}{282}
\begin{document}
\def\logg{$\log (g)$~}
\def\Teff{$T_{\rm eff}$~}
\def\FeH{$[Fe/H]$~}
\def\vmicc{$\xi_{t}$}
\def\vmic{$\xi_{t}$~}
\def\msun{$M_{\odot}$~}
\newcommand{\iso}[2]{{}^{#2}\mbox{#1}}

\title{The s-process nucleosynthesis: 
impact of the uncertainties in the nuclear physics
determined by Monte Carlo variations}

\author {G. Cescutti \inst{1,8} \thanks {email to: cescutti@oats.inaf.it} 
\and N. Nishimura \inst{2,8} \and
  R. Hirschi \inst{3,4,8} \and T. Rauscher \inst{5,6,8} \and 
J. W. den Hartogh \inst{3,8} \and A. St. J. Murphy \inst{7,8}}

\institute{
INAF, Osservatorio Astronomico di Trieste, via G.B. Tiepolo 11,
I-34131 Trieste, Italy
\and Yukawa Institute for Theoretical Physics, Kyoto University, Kyoto 606-8502, Japan
\and Astrophysics Group, Keele University, Keele ST5 5BG, UK
\and Kavli IPMU (WPI), University of Tokyo, Kashiwa 277-8583, Japan
\and Department of Physics, University of Basel, 4056 Basel,
Switzerland
\and Centre for Astrophysical Research, University of Hertfordshire, Hatfield AL10 9AB, UK
\and School of Physics and Astronomy, University of Edinburgh, Edinburgh EH9 3FD, UK
\and BRIDGCE UK Network, www.bridgce.ac.uk, UK}

\authorrunning{Cescutti et al. }

\titlerunning{The s-process nucleosynthesis}

\abstract{We investigated the impact of uncertainties in
  neutron-capture and weak reactions (on heavy elements) on the
  s-process nucleosynthesis in low-mass stars and massive stars using
  a Monte-Carlo based approach.  We performed extensive nuclear
  reaction network calculations that include newly evaluated
  temperature-dependent upper and lower limits for the individual
  reaction rates.  We found $\beta$-decay rate uncertainties affect
  only a few nuclei near s-process branchings, whereas most of the
  uncertainty in the final abundances is caused by uncertainties in
  the neutron capture rates. We suggest a list of uncertain rates as
  candidates for improved measurement by future experiments. }

\maketitle{}

\section{Introduction}

The s-process nucleosynthesis is a source of heavy elements beyond
iron in the universe, taking place in stellar burning
environments. There are two astronomical conditions and corresponding
classes of the s-process \citep[for a review, see][]{2011RvMP...83..157K} and
references therein). The s-process occurs in (i) thermal pulses of low
mass AGB stars producing heavy nuclei up to Pb and Bi, called the {\it
  main} s-process; (ii) He-core and C-shell burning phases of massive stars
representing the lighter components (up to $A \approx 90$), categorised as
the {\it weak} s-process.

In both cases, the primary mechanism is to produce heavier elements
due to the neutron capture and $\beta$-decay close to valley of stability
from seed Fe nuclei over a long-term stellar evolution period. Neutron
source reactions for the s-process are $\alpha$-captures on different
nuclei, where $^{13}{\rm C}(\alpha,{\rm n})^{16}{\rm O}$ and
$^{22}{\rm Ne}(\alpha,{\rm n})^{25}{\rm Mg}$ are main reactions for
the main and weak s-processes, respectively. The impact of these key
fusion reactions has already been studied 
\citep{2011RvMP...83..157K}. The remaining problem is the effects
of uncertainty of (n,$\gamma$) and $\beta$-decay reactions on the final
products. As many of these reactions are involved in the s-process, the
uncertainty is not as simple as the cases of neutron source/poison
reactions. More systematic studies based on the Monte-Carlo (MC) and
statistical analysis \citep{2015JPhG...42c4007I, 2016MNRAS.463.4153R}
are necessary for such problems.

\begin{figure*}[ht!]
\begin{centering}
\vspace{-1cm}

\begin{minipage}{180mm}
\hspace{-15mm}
\includegraphics[width=175mm]{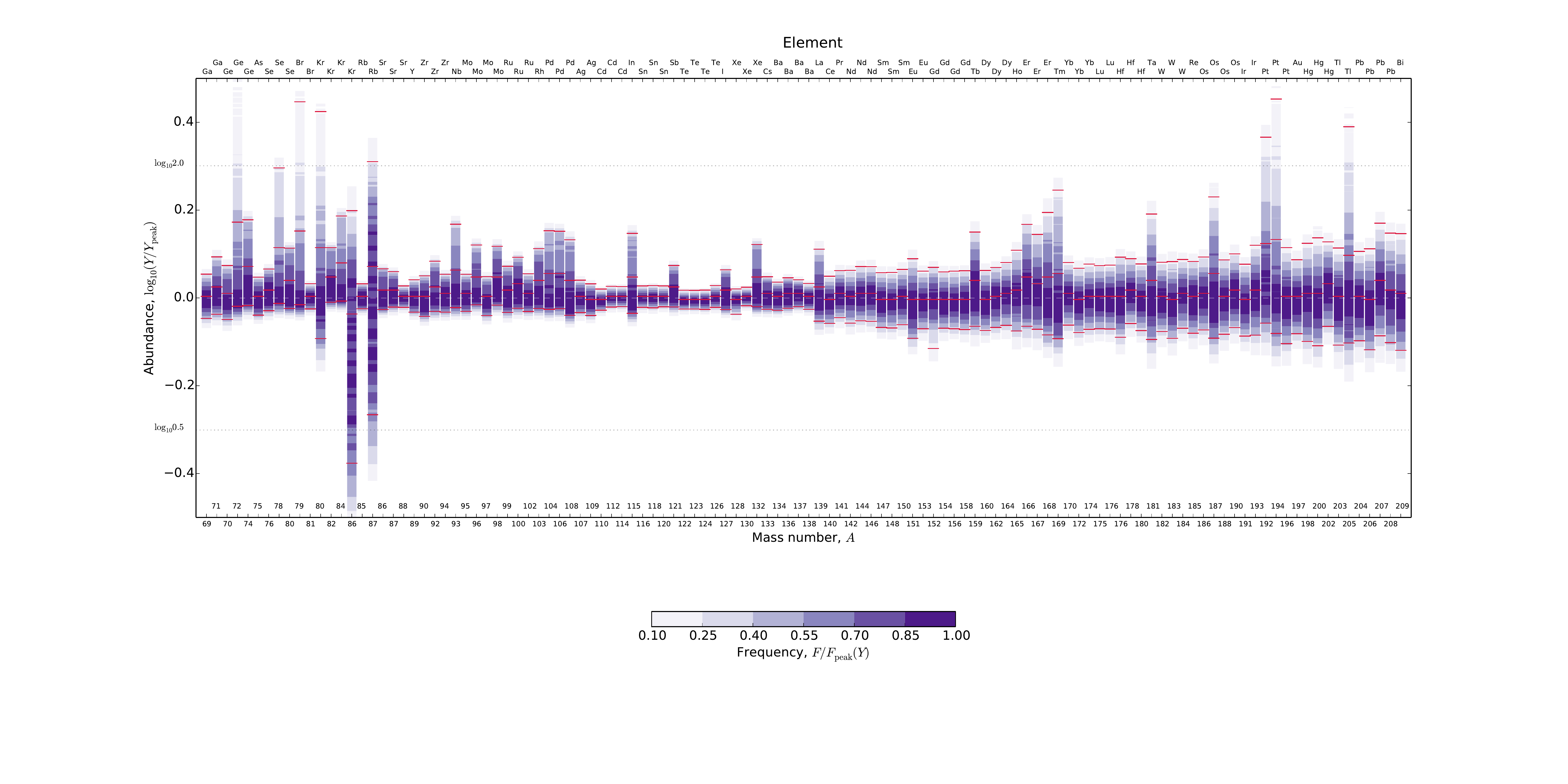}
\end{minipage}
\begin{minipage}{140mm}

  \caption{The results of the MC for the main s-process. Uncertainty
    range is shown for each isotope with the shading and red lines,
    which indicate 5\% and 95\% of the distribution. Note that the
    final abundance is normalised by the value at the peak of the
    distribution.}
\label{fig-mc-ms}
\end{minipage}
\end{centering}
\end{figure*}
In this study, we investigate the impact of uncertainty due to nuclear
physics on the s-process using the MC-based nuclear reaction network. 
Adopting simplified stellar models that
reproduce typical s-process patterns, we apply realistic
temperature-dependent uncertainty of nuclear reaction and decay rates
to nucleosynthesis calculation. Based on an MC method, we evaluate
uncertainty of nucleosynthesis yields.

\section{Methods}

We use simplified stellar evolution models at the solar metallicity
based on 1D evolution calculation. We follow nucleosynthesis evolution
along temporal history of the temperature and density from the initial
abundances. The thermal evolution is treated as the time evolution for
a ``trajectory'' as a single fluid component. We adopt $3M_\odot$ AGB
star model calculated by the MESA code \citep{2011ApJS..192....3P} and
$25 M_\odot$ massive star evolution model
\citep{2004A&A...425..649H,2008IAUS..255..297H}. We confirmed that
these trajectories reproduce a typical abundance pattern for the main
and weak s-process, respectively.

We consider that reaction rates have a temperature-dependent
uncertainty due to the relative contributions by the ground state and
excited states for experimental based cross sections. Following the
prescription in \citet{2011ApJ...738..143R} and \citet{2012ApJS..201...26R},
experimental uncertainties are used for the ground state contributions
to (n,$\gamma$) rates, whereas a factor $2$ is used for excited state
uncertainties \citep[for details, see][]{2012ApJS..201...26R}. As
theoretical calculated rates may have large uncertainty, we simply
apply a constant value $2$.

A similar approach is used for $\beta$-decay rates, based on partition
functions to consider excited state contribution. The uncertainty at
lower temperatures ($T < 10^7$~K) corresponds to the ground state
value, while the uncertainty becomes larger as the temperature
increases. We adopt a factor $10$ for the maximum value at a high
temperature, although uncertainty is about $2$ in stellar burning
temperatures.

\section{Results of MC calculations}
\label{sec-1}

\begin{figure}[t]
\centering
\includegraphics[height=1.\hsize]{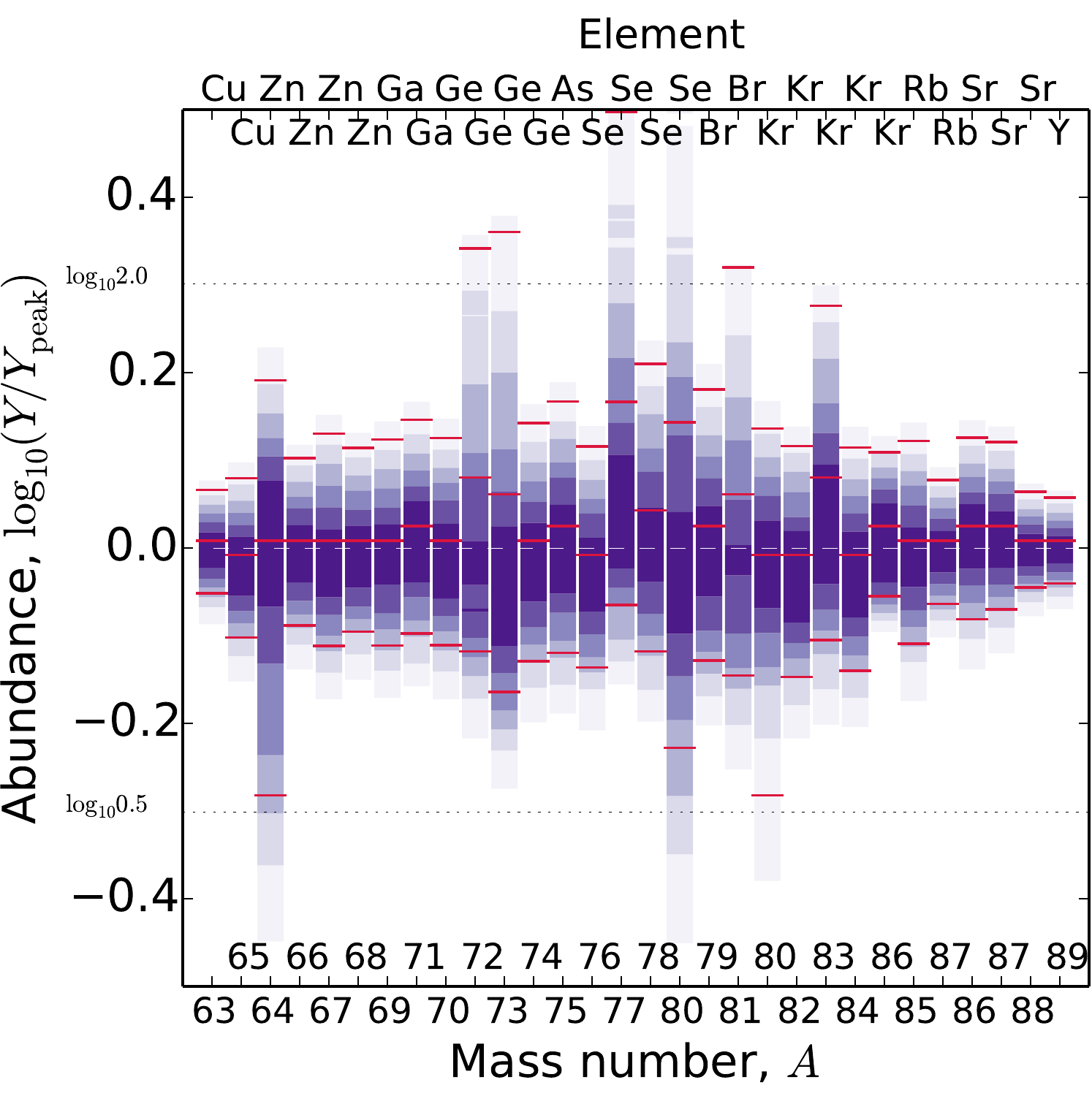}
\caption{The same as Fig.~\ref{fig-mc-ms}, but the results of weak
  s-process. Taken from \citet{Nishimura17}.}
\label{fig-mc-ws}
\end{figure}

We performed MC simulations with variation of reaction rates. A
uniform random distribution between the upper and lower limit of the
reaction rate at a given temperature was used for each variation
. 
Fig.~\ref{fig-mc-ms} shows the resulting production uncertainty of
main s-process for the cases where we varied all (n,$\gamma$)
reactions and $\beta$-decays. 
For the main s-process, we select abundance uncertainties for 116
stable s-process isotopes up to bismuth. Since the number of nuclei
produced by the s-process exceeds 200 species, we only select isotopes
that contribute a minimum of 10\% to the total elemental abundance.
The colour distribution corresponds to the normalised
probability density distribution of the uncertainty in the final
abundance.

Fig.~\ref{fig-mc-ws} shows the resulting production uncertainty of
weak s-process \citep{Nishimura17}. We select in this case abundance uncertainties for
stable s-process isotopes up to $\sim 90$.

The stored MC data allow us for a more comprehensive analsysis and a fully
automated search for key rates. Since the variation factors for
each rate are saved in the stored variation, it can be
tested whether there is a correlation between the variation of a rate
and the resulting change in abundance. The correlation will be larger
the fewer reactions contribute to the uncertainty of the abundance of
a given isotope. There are various definitions for correlations in the
literature. We employ a widely-used correlation coefficient, the
Pearson product-moment correlation coefficient \citep{pearson}.

Review of the available literature suggests that a Pearson
product-moment correlation coefficient value above 0.7 indicates a
strong correlation. We choose a threshold of 0.65 to account for
numerical uncertainties in our calculations.
Therefore we define key rates,  rates with a value above 0.65 
in our MC run.

In Table \ref{tab1}, we compare the correlation coefficient for the
key reactions for the main s-process which are also relevant
for the weak s-process. Not all of the latter are above the
established threshold, but we decide to show them to
have a comparison.

There are several differences, but we underline that for example
$\iso{Ge}{72}(\mbox{n},\gamma)\iso{Ge}{73}$,
$\iso{Se}{78}(\mbox{n},\gamma)\iso{Se}{79}$ and
$\iso{Kr}{85}(\mbox{n},\gamma)\iso{Kr}{86}$ are key rates with very
high correlations, therefore a more precise measurements of these
rates can provide better nucleosynthesis for both processes.

\begin{table*}[]

\begin{minipage}{120mm}
\centering
\caption{First column: key rates determining the production uncertainties for the
    main s-process, and also important for the weak s-process;
 second column: isotopes for which the rate is highly correlated in the
 main s-process production; third column: value of this correlation; 
fourth column: isotopes for which the rate is correlated in the weak 
s-process production; fifth column: value of this correlation.}

\begin{tabular}{ccrcr}
\hline
Key rates & Nuclide& $r_{{\rm cor},0}$ & Nuclide&  $r_{{\rm cor},0}$ \\
             & main s- &main s-         & weak s- &    weak s-          \\
\hline
$\iso{Ge}{72}(\mbox{n},\gamma)\iso{Ge}{73}$ &$\iso{Ge}{72}$ & -0.93 & $\iso{Ge}{72}$ & -0.85 \\
$\iso{Ge}{74}(\mbox{n},\gamma)\iso{Ge}{75}$ &$\iso{Ge}{74}$ & -0.97 & $\iso{Ge}{74}$  & -0.44 \\
$\iso{As}{75}(\mbox{n},\gamma)\iso{As}{76}$ &$\iso{As}{75}$ & -0.86  & $\iso{As}{75}$ & -0.50 \\
$\iso{Se}{78}(\mbox{n},\gamma)\iso{Se}{79}$ & $\iso{Se}{78}$ & -0.96  & $\iso{Se}{78}$ & -0.71 \\
$\iso{Kr}{84}(\mbox{n},\gamma)\iso{Kr}{85}$ & $\iso{Kr}{84}$ & -0.99 &$\iso{Kr}{84}$ &  -0.49\\
$\iso{Kr}{85}(\mbox{n},\gamma)\iso{Kr}{86}$ & $\iso{Kr}{86}$ & 0.88   &$\iso{Kr}{86}$  &  0.84  \\
\hline
\end{tabular}\label{tab1}
\end{minipage}

\end{table*}

\section{Conclusion}

We evaluated the impact on s-process nucleosynthesis in massive stars
and low mass AGB stars of nuclear physics uncertainties in neutron
capture and weak reactions on heavy elements using MC
calculations. Our method is a robust way to identify key reaction
rates to support further investigations in nuclear astrophysics
regarding the s-process. The method can identify the importance of
reactions and we found that
$\iso{Ge}{72}(\mbox{n},\gamma)\iso{Ge}{73}$,
$\iso{Se}{78}(\mbox{n},\gamma)\iso{Se}{79}$ and
$\iso{Kr}{85}(\mbox{n},\gamma)\iso{Kr}{86}$ are key rates with very
high correlations for both weak s-process and main s-process.  More
detailed analysis are presented in \citet{Nishimura17} for s-process
in massive stars; whereas for the main s-process, they will be shown
in our upcoming \citet{Cescutti17}.

\begin{acknowledgements}
This project has been financially supported by the ERC
(EU-FP7-ERC-2012-St Grant 306901, EU-FP7 Adv Grant GA321263-FISH) and
the UK STFC (ST/M000958/1). Parts of computations were carried out by
COSMOS (STFC DiRAC Facility) at DAMTP in University of Cambridge.
G.C. acknowledges financial support
from the European Union Horizon 2020 research and
innovation programme under the Marie Sk\l odowska-Curie
grant agreement No. 664931.
\end{acknowledgements}

\bibliographystyle{aa}
\bibliography{Nobu}

\begin{thebibliography}{11}
\expandafter\ifx\csname natexlab\endcsname\relax\def\natexlab#1{#1}\fi

\bibitem[{{Cescutti} {et~al.}(2017){Cescutti}, {Nishimura}, {Hirschi},
  {Rauscher}, {den Hartogh}, \& {Murphy}}]{Cescutti17}
{Cescutti}, G., {Nishimura}, N., {Hirschi}, R., {et~al.} 2017, \mnras \ in
  preparation

\bibitem[{{Hirschi} {et~al.}(2008){Hirschi}, {Frischknecht}, {Pignatari},
  {Thielemann}, {Bennet}, {Diehl}, {Fryer}, {Herwig}, {Hungerford},
  {Magkotsios}, {Rockefeller}, {Timmes}, \& {Young}}]{2008IAUS..255..297H}
{Hirschi}, R., {Frischknecht}, U., {Pignatari}, M., {et~al.} 2008, in Nuclei in
  the Cosmos (NIC X)

\bibitem[{Hirschi {et~al.}(2004)Hirschi, Meynet, \&
  Maeder}]{2004A&A...425..649H}
Hirschi, R., Meynet, G., \& Maeder, A. 2004, Astronomy and Astrophysics, 425,
  649

\bibitem[{Iliadis {et~al.}(2015)Iliadis, Longland, Coc, Timmes, \&
  Champagne}]{2015JPhG...42c4007I}
Iliadis, C., Longland, R., Coc, A., Timmes, F.~X., \& Champagne, A.~E. 2015,
  Journal of Physics G Nuclear Physics, 42, 034007

\bibitem[{K\"appeler {et~al.}(2011)K\"appeler, Gallino, Bisterzo, \&
  Aoki}]{2011RvMP...83..157K}
K\"appeler, F., Gallino, R., Bisterzo, S., \& Aoki, W. 2011, Reviews of Modern
  Physics, 83, 157

\bibitem[{{Nishimura} {et~al.}(2017){Nishimura}, {Hirschi}, {Rauscher},
  {Murphy}, \& {Cescutti}}]{Nishimura17}
{Nishimura}, N., {Hirschi}, R., {Rauscher}, T., {Murphy}, A.~S.~J., \&
  {Cescutti}, G. 2017, \mnras, 469, 1752

\bibitem[{{Paxton} {et~al.}(2011){Paxton}, {Bildsten}, {Dotter}, {Herwig},
  {Lesaffre}, \& {Timmes}}]{2011ApJS..192....3P}
{Paxton}, B., {Bildsten}, L., {Dotter}, A., {et~al.} 2011, \apjs, 192, 3

\bibitem[{Pearson(1895)}]{pearson}
Pearson, K. 1895, Proceedings of the Royal Society of London Series I, 58, 240

\bibitem[{Rauscher(2012)}]{2012ApJS..201...26R}
Rauscher, T. 2012, The Astrophysical Journal Supplement Series, 201, 26

\bibitem[{Rauscher {et~al.}(2011)Rauscher, Mohr, Dillmann, \&
  Plag}]{2011ApJ...738..143R}
Rauscher, T., Mohr, P., Dillmann, I., \& Plag, R. 2011, The Astrophysical
  Journal, 738, 143

\bibitem[{Rauscher {et~al.}(2016)Rauscher, Nishimura, Hirschi, Cescutti,
  Murphy, \& Heger}]{2016MNRAS.463.4153R}
Rauscher, T., Nishimura, N., Hirschi, R., {et~al.} 2016, Monthly Notices of the
  Royal Astronomical Society, 463, 4153

\end{thebibliography}

\end{document}